\documentstyle[prl,aps,twocolumn]{revtex}
\begin{document}

\def\del {\nabla}
\def\vq  {{\bf q}}
\def\vk  {{\bf k}}
\def\vx  {{\bf x}}

\title{Nonlinearities in Conservative Growth Equations}
\author{Abhijit K. Kshirsagar}
\address{Raman Research Institute, Bangalore 560080, India}
\author{S. V. Ghaisas}
\address{Department of Electronics Science, University of Pune,
Pune 411007, India}

\maketitle

\begin{abstract}
Using the dynamic renormalization group (DRG) technique, we analyze general
nonlinearities in a conservative nonlinear growth equation with
non-conserved gaussian white noise. We show that they fall in two classes
only: the Edwards-Wilkinson and Lai-Das Sarma types, by explicitly computing
the associated amputated two and three point functions at the first order in
perturbation parameter(s). We further generalize this analysis to higher
order nonlinearities and also suggest a physically meaningful geometric
interpretation of the same.
\end{abstract}

Growth of thin films from vapor has been a subject of significant interest
for both experimentalists and theorists. The non-equilibrium nature of this
growth makes the techniques applicable to the study of fractals, available for
understanding the physics of thin film growth; particularly in the
thermodynamic
limit \cite{bar}. Most of the characterization can be done by observing
 spatio temporal behavior of the interface fluctuations.
This behavior can be studied within the framework of continuum
equations such as a Langevin type
equation proposed by Edwards and Wilkinson (EW) \cite{EW}. It allows
a non-equilibrium statistical mechanical description which brings out the
scale invariant nature of the interface roughness. Consequently one obtains two
dynamical critical exponents $\alpha$ and $\beta$ (or equivalently $\alpha$
and $z=\alpha/\beta$) which completely characterize the growth. The
correlation length along the substrate varies as $r^\alpha$ and
$t^{1/z}$ whereas that along the normal to the surface exhibits a $t^\beta$
behavior.

The geometrical feature of growth represented by the lateral component of
the growth velocity was captured by Kardar, Parisi and Zhang (KPZ)\cite{KPZ},
by introducing a nonlinear term proportional to $(\del h)^2$ in the EW
equation:
\begin{equation}
{ \partial h \over \partial t} = \nu \del^2 h + { \lambda \over 2}
\big( \del h \big)^2 + \eta
\end{equation}
where $h(x,t)$ is the height function and $\eta (x,t)$ is a non-conserved
gaussian white noise characterized completely by its one and two-point
functions:
\begin{eqnarray}
\langle \eta (\vx , t) \rangle &=& 0 \nonumber \\
\mbox{and} \hskip 0.25in
\langle \eta (\vx ,t) \eta (\vx^{\prime} , t^{\prime}) \rangle & = &
2 D \delta (\vx - \vx^{\prime}) \delta (t - t^{\prime})
\end{eqnarray}

Dynamic renormalization group technique \cite{FNS} applied to the KPZ
model yields
$z=3/2$ and $\alpha=1/2$, based on one-loop calculations for a one
dimensional substrate. These values of $\alpha$ and $z$ are obtained by
looking for a nontrivial fixed point for the set of one-loop flow
equations for the parameters $\nu$, $\lambda$ and $D$ stated above.
It turns out that these results are exact in the sense that they hold to all
orders and can be obtained by invoking the Galilean invariance of the KPZ
equation and the Fokker-Planck equation associated with it for the $1+1$ case
\cite{bar}.
The KPZ equation, however, is of the {\it non-conservative} type, (i.e. it
can {\it not} be written in the form ${\partial h\over {\partial t}}=
\del\cdot j$ where $j({\bf x},t)$ is a local current) and so is relevant mainly
for growth with overhangs and vacancies corresponding to the regime of
high temperature and high volatility of the deposited material.

A general fourth order {\it conservative} equation was proposed by Lai
and Das Sarma (LDS)\cite{LDS} of the form
%
%
\begin{eqnarray}
{ \partial h \over \partial t} = \nu_2 \del^2 h &-&  \nu_4 \del^4 h
+ { \lambda_{22} } \del^2 \big( \del h \big)^2 \nonumber \\
&+& \lambda_{13} \del \cdot \big( \del h \big)^3 + \eta
\end{eqnarray}
If $\lambda_{22}=0$ then in the absence of noise, this equation is
symmetric under $h \rightarrow -h$. When $\nu_2$ is zero
 the $\del \cdot (\del h)^3$ term is more relevant than the
$\del^2 {(\del h)}^2$ term as a perturbation to the linear term
$\del^4 h$. If one makes the coefficients in this equation time dependent,
the situation can change significantly with the possibility of the latter
term taking over the former one. Interpretation of this in terms of
different relaxation mechanisms operating at different times during the
pre-asymptotic region in time will be discussed elsewhere\cite{GKD}.

In a recent paper, Das Sarma and Kotlyar \cite{DSK}, have shown that the
most general conserved fourth order nonlinearity, viz.
$\lambda_{13}\del\cdot (\del h)^3$ belongs to the universality class
of the EW model., i.e. the interaction produces a $-\nu_2 k^2$ term in
the inverse propagator at zero frequency, where the coefficient $\nu_2$
is proportional to the {\it first} power of the perturbation parameter
$\lambda_{13}$. This confirmed the earlier numerical evidence based on
simulations in $(1+1)$ dimensions due to Kim and Das Sarma\cite{KDS}.

The purpose of this letter is to show that this analysis is in fact
completely general and a general nonlinear conservative growth equation
reduces to either Lai-Das Sarma or the EW form. We shall consider the
following equation with a general nonlinearity of order$(2n+2)$ for
integer $n$,
\begin{eqnarray}
{ \partial h \over \partial t} = \nu \del^{2n+2} h
&+& { \lambda_{2} \over 2n } \del^2 \big( \del h \big)^{2n}  \nonumber \\
&+& { \lambda_{1}
\over (2n + 1) } \del \cdot \big( \del h \big)^{2n+1}
+ \eta
\end{eqnarray}
Note that the linear term too is of order $(2n+2)$, in order to keep
the nonlinear terms relevant in RG sense.

Consider e.g., the general nonlinearity:~$\del \cdot (\del h)^{2n+1}$.
(The
choices $n=0$ and $n=1$ correspond respectively to the EW equation and
reference [6]). For the purpose of diagrammatic perturbation theory, it
corresponds to a ``vertex" with $(2n+1)$ prongs ending on a cross and
one free ended prong.(see, e.g., the book by Barab\'asi and Stanley
for notations and conventions).
The vertex and the corresponding Feynman rule are in Fig.1.
\linebreak
for example $n=1$ it corresponds to the vertex factor of
${1\over 3} \lambda_1 (\vq_1\cdot \vq_2) \vk \cdot (\vk-\vq_1-\vq_2)$,
which upon suitable identification of the parameters and a change of variables
becomes the expression of reference [6]. The propagator modification at
order $\lambda_1$ is shown in Fig.2.

Again, with $n=1$ we recover the one-loop {\it Hartree} term of reference [6].
The $n$-loop Hartree diagram shown in figure 2, is equally easy to
evaluate; thanks
to the peculiar nature of the noise contraction which generates the
Feynman factor $2D \delta(\vk_1+\vk_2) \delta(\omega_1+\omega_2)$. The
inverse free propagator is given by
\begin{equation}
G_0^{-1} (\vk , \omega) = - i \omega + (-1)^n \nu k^{2n + 2}
\end{equation}
For our purpose (which is to identify the modification of the equation of
motion or the hamiltonian whose $h$-variation produces
the equation), it is enough to compute the `vertex function' or the
`amputated' correlation function which amounts to snapping off the
external propagators. There is a symmetry factor associated with this diagram
given by ${1\over 2}(2n+1)\cdot 2n(2n-1)$. Also every noise contraction
gives a $2D$ factor yielding an overall multiplier as
$(-1)^{n+1} \lambda_1 \cdot n(2n-1)(2D)^n$. Using the standard techniques to
evaluate the frequency-momentum integrals, we finally arrive at the
amputated $n$-loop Hartree vertex expression as
\begin{equation}
(-1)^{n+1} \lambda_1 (D/\nu)^n n (2n-1) K_d^n
\Big[ \int q^{d-(2n+1)}dq \Big]^n \cdot k^2,
\end{equation}
where $K_d$ is related to the usual angular integration factor as
$K_d={1 \over {(2\pi)}^d}S_d$.
It is clear from this expression that the $\del \cdot (\del h)^{2n+1}$
nonlinearity produces a $\del^2 h$ term (identified as the coefficient
of $-k^2$ in the above expression) at the {\it same} order in $\lambda_1$
in dimensions $d < d_{crit}$. This puts the ``odd order"
nonlinearities in the EW class.

Exactly similar analysis (see Fig.3 )
of the ``even order" nonlinearity
${1\over {2n}}\lambda_2 \del^2(\del h)^{2n}$ produces the
{\it amputated three point function} (which is just the object that
appears in the equation) given by
\begin{equation}
k^2 \big[ \vq ( \vk - \vq ) \big]  { 3 \lambda_2 \over n } { 2n \choose 4}
\Big( {D K_d \over \nu} \Big)^{n-1} \Big[ \int p^{d - (2n+1) } dp \Big]^{n-1}
\end{equation}
Again, we simply identify from this expression that the Lai-Das Sarma
term, $\del^2 (\del h)^2$ is produced at the {\it first
order} in $\lambda_2$ in dimensions less than the critical dimension,
putting the even order nonlinearities in the Lai-Das Sarma class.

These arguments are trivially extendible to the nonlinearities of the
form $\del^{2m+1} \cdot (\del h)^{2n+1}$ and $\del^{2m} (\del h)^{2n}$
(with the appropriate linear terms in the growth equation to keep
the nonlinearities relevant). They respectively generate the
generalized EW and the Lai-Das Sarma terms, viz. $\del^{2m} \del^2 h$
and $\del ^{2m} (\del h)^2$ types. Generally a nonlinearity of the
kind $ \del^2 [ \del^{2m} h ]^{2n}$ collapses to $\del^2  (\del^{2m} h)^2$
and $ \del \cdot [ \del^{2m+1} h ]^{2n+1}$ to $\del^2 [\del^{2m} h]$.

The above mechanism of generation of the lower order terms via higher
order nonlinearities reflects onto a simple physically meaningful
geometric picture of the evolving growth front.
Consider e.g., the equation
\begin{equation}
{ \partial h \over \partial t} = \nu \del^{2n+2} h
+  \lambda_{1}
\del \cdot \big( \del h \big)^{2n+1}
+ \eta
\end{equation}

The rate ${\partial h\over \partial t}$ depends upon the variation of
$(\del h)^{2n+1}$, including its sign
 and $h$ varies locally
to attempt to make this rate vanish for a stable growth mode.
Note that the nature of the response
for terms $\del \cdot (\del h)$ and $\del \cdot (\del h)^{2n+1}$ is exactly the
same
except for its magnitude, given {\it any geometrical configuration on the
surface} for any value of $n$ (i.e. the power of the gradient).
This fact seems, according to us, to be the root cause for the
generation of the EW term since the surface morphology in these cases
must be comparable, considering the compatibility of the responses.

Similar reasoning can be furnished in the case of $\del ^2(\del h)^{2n}$
leading to the $\del^2 (\del h)^2$ term. Here the only difference is that the
growth rate does {\it not} depend on the sign of $h$, thus making the
LDS term as the lowest order term allowed. This leads to the conclusion
that in general the nonlinearities are characterized by the order of the
variation of $h$ locally, i.e. slope, curvature, etc..., corresponding
to $\del h, \del^2 h, \cdots$ etc.. and thus a general nonlinearity is
of the form $\del^{2m} (\del^n h)^2$.

\acknowledgements
The authors would like to thank Professor S. Das Sarma for a critical
reading of the manuscript, valuable suggestions and many
insightful remarks which contributed significantly to the
improvement of this manuscript.
One of us (AKK) would also like to thank Professor Martin Grant for useful
correspondence and the Department of Physics, University of Pune for
hospitality during a long term visit in which the above work was
completed.

\noindent Electronic Mail:

abhi@rri.ernet.in, abhi@iucaa.ernet.in (AKK),

elec@physics.unipune.ernet.in (SVG),

\begin{figure}
\caption{The vertex for $ \nabla \cdot \big(\nabla h \big)^{2n+1}$
nonlinearity. The associated Feynman rule is: }
$ { (-1)^{n+1} \over (2n+1) } \lambda_1
\big( {\bf q}_1 \cdot {\bf q}_2 \big)
\cdots
 \big( {\bf q}_{2n-1} \cdot {\bf q}_{2n} \big)
\big[{\bf k} \cdot {\bf q}_{2n+1} \big]
\delta ( {\bf k} - \sum_{i=1}^{2n+1} {\bf q}_i)
\delta ( \omega - \sum_{i=1}^{2n+1} \omega_i) $
\label{Fig1}
\end{figure}

\begin{figure}
\caption{ The $n$-Loop Hartree diagram generating the EW term with
a coefficient proportional to $\lambda_1$. Also shown is the two-point
vertex function.}
\label{Fig2}
\end{figure}

\begin{figure}
\caption{ The diagram for the nonlinearity of the type $ \nabla^2
(\nabla h )^2 $ proportional to $\lambda_2$ generated by the
nonlinearity $ {1 \over 2n} \lambda_2 \nabla^2 ( \nabla h)^{2n}$.
Also shown is the three-point vertex function.}
\label{Fig3}
\end{figure}

\end{document}